\documentclass[prl,aps,reprint,twocolumn,citeautoscript,groupedaddress,noeprint,nofootinbib]{revtex4-2}

\usepackage{stix}
\usepackage{amsmath}
\usepackage{bm}
\usepackage{graphicx,color}
\usepackage{subfigure}
\usepackage{array}
\usepackage{braket}
\usepackage{nicefrac}
\usepackage[dvipsnames]{xcolor}

\usepackage{booktabs}

\usepackage{color}
\definecolor{myblue}{rgb}{0,0,1}
\usepackage[breaklinks=true,colorlinks=true,linkcolor=myblue,urlcolor=myblue,citecolor=myblue]{hyperref}

\usepackage{verbatim}

\newcommand*{\gw}{{\textit{G}\textsubscript{0}\textit{W}\textsubscript{0}}}
\newcommand*{\GW}{\textit{GW}}

\begin{document}

\title{Interacting-bath dynamical embedding for capturing non-local electron correlation in solids}

\author{Jiachen Li}
\author{Tianyu Zhu}
\email{tianyu.zhu@yale.edu}
\affiliation{Department of Chemistry, Yale University, New Haven, CT 06520, USA}

\begin{abstract}
Quantitative simulation of electronic structure of solids requires treating local and non-local electron correlations on an equal footing. We present a new \textit{ab initio} formulation of Green's function embedding which, unlike dynamical mean-field theory that uses non-interacting bath, derives bath representation with general two-particle interactions in a systematically improvable manner. The resulting interacting-bath dynamical embedding theory (ibDET) utilizes an efficient real-axis coupled-cluster solver to compute the self-energy, approaching the full system limit at much reduced cost. When combined with the \textit{GW} theory, \textit{GW}+ibDET achieves good agreement with experimental spectral properties across a range of semiconducting, insulating and metallic materials. Our approach also enables quantifying the role of non-local electron correlation in determining material properties and addressing the long-standing debate on the bandwidth narrowing of metallic sodium.
\end{abstract}

\maketitle

\newpage

\textit{Introduction.} Predictive description of material-specific electronic properties 
remains a significant challenge in computational physics and chemistry~\cite{Kent2018a}. The main reason is the need for quantitative treatment of electron correlation effects and simulating in the thermodynamic limit simultaneously. Quantum embedding theories offer a promising route to solve this problem~\cite{Sun2016}. 
For dynamical quantities, dynamical mean-field theory (DMFT) has been the most popular choice, leading to advances in the understanding of correlated electron physics in lattice models and real materials~\cite{georgesHubbardModelInfinite1992,georgesDynamicalMeanfieldTheory1996,Kotliar2006}.

Despite its success in treating strong local electron interactions, extending DMFT to accurately capture non-local electron correlation remains challenging~\cite{Rohringer2018a}. This capability is crucial for describing various quantum many-body phenomena, such as the pseudogap phase and stripe orders in high-temperature cuprate superconductors~\cite{hashimotoEnergyGapsHightransitiontemperature2014,Zheng2017,Xu2024a}. Cluster extensions in real (cluster DMFT~\cite{Lichtenstein2000,kotliarCellularDynamicalMean2001,biroliClusterMethodsStrongly2002}) or reciprocal (DCA and D$\Gamma$A~\cite{Hettler1998,maierQuantumClusterTheories2005,Toschi2007}) spaces have been proposed, but these formalisms are mostly designed for short-range quantum fluctuations, and CDMFT is known to break translational invariance~\cite{Klett2020a}. To account for band structure and long-range interactions in real materials, density functional theory (DFT)~\cite{kohnSelfConsistentEquationsIncluding1965} or many-body perturbation theory (\GW)~\cite{Hybertsen1986,Golze2019a,zhuAllElectronGaussianBasedG0W02021,leiGaussianbasedQuasiparticleSelfconsistent2022} is normally adopted as the low-level theory for DMFT. Although much progress has been made in the downfolded DFT+DMFT~\cite{Kotliar2006} and \GW+DMFT~\cite{Sun2002,Biermann2003,nilssonMultitierSelfconsistentGW2017,Tomczak2017} formalisms, 
their predictive capability is limited by uncontrolled errors that are often difficult to quantify. The impurity problem usually comprises a few correlated orbitals, but DMFT results could depend sensitively on the choice and construction of these impurity orbitals~\cite{Karp2021a}. The derivation of effective interactions and approximation to their frequency dependence also introduce numerical uncertainties~\cite{Aryasetiawan2004}. Moreover, DFT+DMFT calculations could suffer from the double counting error~\cite{Wang2012,Muechler2022b}.

To avoid these numerical ambiguities, one of us recently developed a full cell \GW+DMFT formalism ~\cite{zhuInitioFullCell2021,zhuEfficientFormulationInitio2020,Zhu2024Kondo}, 
where the impurity problem comprises all local orbitals of atoms within a chosen supercell. General bare Coulomb interactions within impurity orbitals are employed and solved by efficient quantum chemistry solvers~\cite{zhuCoupledclusterImpuritySolvers2019,Ronca2017a}, removing the need for downfolding. 
However, full cell \GW+DMFT inevitably inherits certain limitations from cluster DMFT, such as the breaking of translational invariance. While the impurity space is significantly larger, the non-local electron correlation beyond the selected supercell is at best captured at the \GW~level. The neglect of long-range interactions stronger than those captured by \GW~is known to yield errors in a variety of settings~\cite{Kutepov2021,nilssonMultitierSelfconsistentGW2017}.

A common origin of this challenge in DMFT is the \textit{non-interacting} nature of its bath representation through the hybridization function. Despite a natural choice for continuous-time quantum Monte Carlo (CTQMC) solvers~\cite{Gull2011}, the non-interacting bath parametrization does not fully leverage the power of Hamiltonian-based solvers, such as exact diagonalization (ED)~\cite{Liebsch2012}, density matrix renormalization group (DMRG)~\cite{Ronca2017a,Zhai2023a}, configuration interaction (CI)~\cite{Zgid2012}, and coupled-cluster (CC) theory~\cite{Nooijen1993,zhuCoupledclusterImpuritySolvers2019,Peng2018c,sheeCoupledClusterImpurity2019}, as there is no clear mapping between the full Hamiltonian and fictitious bath states. In this Letter, we develop a new \textit{ab initio} Green's function embedding formulation with \textit{interacting} bath, which enables direct projection of the full interacting Hamiltonian into large embedding problems,
solved by a coupled-cluster Green's function (CCGF) solver truncated at the single-reference singles and doubles level~\cite{zhuCoupledclusterImpuritySolvers2019}. 
Unlike DMFT, this formulation utilizes self-energy corrections to both impurity and bath states for describing dynamical quantities, allowing the computation of coupled-cluster spectra at substantially reduced cost. Because we do not derive bath parameters by fitting the hybridization function, this method is, strictly speaking, no longer DMFT, and we term it interacting-bath dynamical embedding theory (ibDET). 

\begin{figure}
\includegraphics[width=0.47\textwidth]{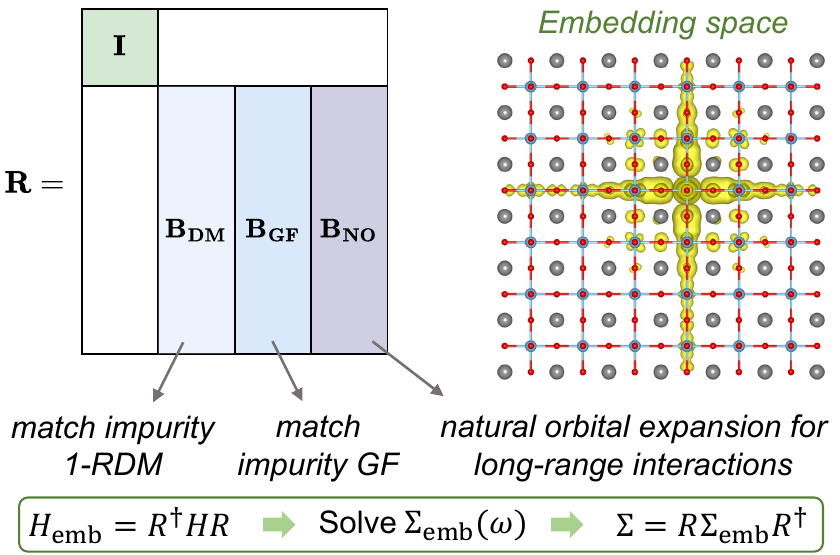}
\caption{Illustration of the ibDET formalism, where each embedding space includes all local orbitals of an impurity atom (``$I$'') in the unit cell coupled to large interacting bath (``$B$''). An example of an occupied embedding space in SrTiO\textsubscript{3} (Ti as impurity) is shown.} 
\label{fig:bath_demo}
\end{figure}

\textit{Method}. Given a periodic crystal, we start with a mean-field solution at the Hartree-Fock (HF) or DFT level using crystalline Gaussian atomic orbitals. To define the impurity problem, we construct the
orthogonal atom-centered local orbital basis employing an intrinsic atomic orbital plus projected atomic orbital (IAO+PAO) scheme~\cite{kniziaIntrinsicAtomicOrbitals2013,cuiEfficientImplementationInitio2020}. We choose all local orbitals on a single atom as the impurity and then gradually expand the bath space by selecting orbitals that entangle most strongly with impurity orbitals from the environment. To recover the self-energy of the full crystal, multiple embedding problems need to be formulated, each centered on an impurity atom in the unit cell.  

The key step is then to perform algebraic construction of bath orbitals that allow projection from full Hamiltonian to the embedding space (Fig.~\ref{fig:bath_demo}). Here, we construct bath orbitals responsible for capturing short- and long-range electron correlations respectively.
The \textit{first} set $B_\mathrm{DM}$ is derived by a Schmidt decomposition, i.e., using the singular value decomposition (SVD) of the mean-ﬁeld oﬀ-diagonal one-particle reduced density matrix (1-RDM) between the impurity and remaining lattice~\cite{cuiEfficientImplementationInitio2020}, the same as in density matrix embedding theory~\cite{kniziaDensityMatrixEmbedding2012}. $B_\mathrm{DM}$ ensures that impurity 1-RDM is exactly reproduced in the embedding calculation at the mean-field level~\cite{bath_dm}. The \textit{second} set $B_\mathrm{GF}$ is obtained by performing SVD of the imaginary part of mean-field off-diagonal Green's function $g(\omega_n)$ on a uniform real-frequency grid, to capture the frequency-dependent entanglement between impurity and environment~\cite{nusspickelEfficientCompressionEnvironment2020}, a role similar to that of the hybridization function in DMFT. 
To keep the number of $B_\mathrm{DM}$ and $B_\mathrm{GF}$ orbitals tractable, we couple bath orbitals only to valence impurity orbitals (i.e., IAOs) and adopt an additional projection to orthogonalize the embedding space and remove redundant $B_\mathrm{GF}$ orbitals.
$B_{\text{DM}}$ and $B_{\text{GF}}$, however, 
do not capture electron correlation beyond short range, so we derive a \textit{third} set of cluster-specific natural bath orbitals $B_{\text{NO}}$ inspired by local correlation methods in quantum chemistry, particularly the local natural orbital coupled-cluster (LNO-CC) theory~\cite{Rolik2011}. Similar idea was recently introduced to quantum embedding by Nusspickel and Booth~\cite{nusspickelSystematicImprovabilityQuantum2022} and periodic CC theory~\cite{Ye2024} for ground-state properties. The key is to select natural orbitals from the environment that correlate strongly  
to the existing embedding cluster ($I \bigoplus B_\mathrm{DM} \bigoplus B_\mathrm{GF}$, where $I$ stands for the impurity space), estimated by a cheap direct second-order perturbation theory (dMP2) calculation. 
Furthermore, to better describe the delocalized conduction states in gapped systems, we incorporate a few low-lying canonical virtual orbitals into the bath space~\cite{bath_no}.

The Hamiltonian for each embedding problem is 
\begin{align}
    H_\mathrm{emb} = \sum_{ij}^\mathrm{emb} \tilde{F}_{ij} a_i^{\dagger} a_j + \frac{1}{2} \sum_{ijkl}^\mathrm{emb} (ij|kl) a_i^{\dagger} a_k^{\dagger} a_l a_j 
    \label{eq:Hemb}
\end{align}
where $(ij|kl)$ is the general two-particle bare Coulomb interaction matrix defined on all impurity and bath orbitals, obtained through a projection with rotation matrix $R$ (see Fig.~\ref{fig:bath_demo} for the definition of $R$). The one-particle interaction matrix is defined as
\begin{equation}
    \tilde{F}_{ij} = F^\mathrm{emb}_{ij} - \sum_{kl}^{\mathrm{emb}} \gamma^\mathrm{emb}_{kl} [(ij|lk) - \frac{1}{2} (ik|lj)].
    \label{eq:femb}
\end{equation}
Here, $F^\mathrm{emb} = R^{\dagger} F^\mathrm{full} R$, where $F^\mathrm{full}$ is the Fock matrix of the full system computed using HF (even when we start from the DFT density), and $\gamma^\mathrm{emb}$ is the 1-RDM rotated to the embedding space. The HF contribution to the self-energy is exactly removed in Eq.~\ref{eq:femb}, so there is no double counting in ibDET. 

The CCGF solver at the EOM-CCSD level~\cite{zhuCoupledclusterImpuritySolvers2019} is adopted to solve the embedding Hamiltonians (Eq.~\ref{eq:Hemb}) on the real axis. We choose the CCGF solver because of its good performance for various lattice models and real materials~\cite{mcclainGaussianBasedCoupledClusterTheory2017,laughonPeriodicCoupledClusterGreen2022,Gruber2018,Mihm2021b,Xing2024}, as well as high computational efficiency. Meanwhile, we emphasize that ibDET can utilize any Hamiltonian-based solvers, such as quantum chemistry DMRG~\cite{Zhai2023a} and selected configuration interaction~\cite{Sharma2017} that are more robust for stronger correlation. The self-energy computed within the embedding space is rotated back to the full Hilbert space
\begin{equation}
    \Sigma^{\mathrm{full},J}(\omega) =
    R \Sigma^{\mathrm{emb},J}(\omega) R^{\dagger},
    \label{eq:sigmarotation}
\end{equation}
where $J$ means the $J$-th embedding problem. The self-energy matrices $\{\Sigma^{\mathrm{full},J}(\omega)\}$ from all embedding calculations are then assembled using a democratic partitioning scheme and Fourier transformed to the momentum space, to obtain the full self-energy of the crystal $\Sigma^\mathrm{ibDET}(\mathbf{k}, \omega)$. Similar to \GW+DMFT, ibDET can be easily combined with the \GW~theory to capture any small long-range correlation effects missed by ibDET, and the resulting \GW+ibDET self-energy is
\begin{equation}
    \Sigma^{\GW+\mathrm{ibDET}} = \Sigma^{\GW, \mathrm{full}} + \Sigma^\mathrm{CC, ibDET} - \Sigma^{\GW, \mathrm{ibDET}} ,
\end{equation}
where $\Sigma^{\GW, \mathrm{full}}$ is the \GW~self-energy of the full system. Different from common DFT+DMFT and \GW+DMFT calculations that require self-consistency, all results in this work are obtained from one-shot ibDET.

\textit{Results}.
We first demonstrate numerical convergence of ibDET results on silicon (Si) and two-dimensional hexagonal boron nitride (2D BN), where full EOM-CCSD calculations are possible. For Si and 2D BN, GTH-cc-pVTZ/GTH-DZVP basis sets~\cite{yeCorrelationConsistentGaussianBasis2022,vandevondeleQuickstepFastAccurate2005} and GTH-HF-rev/GTH-PADE pseudopotentials~\cite{hartwigsenRelativisticSeparableDualspace1998,goedeckerSeparableDualspaceGaussian1996} were employed, together with $4 \times 4 \times 4$/$6 \times 6 \times 1$ \textbf{k}-point sampling.
For Si, it is not feasible to run full EOM-CCSD calculation, thus we used a composite correction scheme~\cite{voPerformancePeriodicEOMCCSD2024} to estimate band gaps. 
All calculations were conducted using the PySCF quantum chemistry software package~\cite{Sun2018c,Sun2020b}.

\begin{figure}
\includegraphics[width=0.48\textwidth]{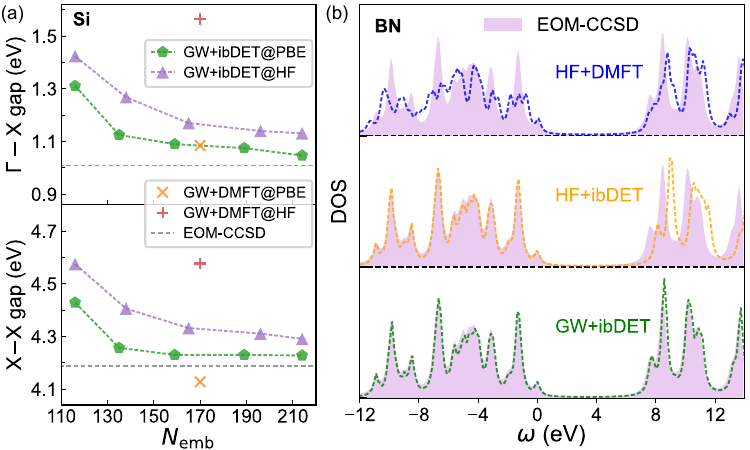}
\caption{Benchmark of ibDET on Si and 2D BN against EOM-CCSD.
(a) Convergence of \GW+ibDET band gaps of Si as the number of embedding orbitals is increased, compared to full cell \GW+DMFT values~\cite{zhuInitioFullCell2021}.
(b) DOS of 2D BN from HF+ibDET and \GW+ibDET (HF reference), compared to the HF+DMFT spectrum~\cite{zhuEfficientFormulationInitio2020}.
}
\label{fig:bn_si}
\end{figure}
\twocolumngrid

In Fig.~\ref{fig:bn_si}a, we show the convergence of \GW+ibDET predictions of silicon band gaps against the full system limit, which is challenging for quantum embedding methods due to the long-range nature of the screened interaction~\cite{zhuEfficientFormulationInitio2020,zhuInitioFullCell2021,Chibani2016a}.
Although one-shot \gw~approximation on top of PBE~\cite{perdewGeneralizedGradientApproximation1996} predicts accurate band structure for Si, this success benefits from error cancellations, indicated by the large difference between \gw@PBE (1.15 eV) and \gw@HF (1.86 eV) $\Gamma$-X band gaps. The \GW+ibDET predicted band gaps quickly converge to the full EOM-CCSD limit as the embedding space grows. At around 210 embedding orbitals (6\% of the total number of orbitals, $N_\mathrm{tot}$), the \GW+ibDET (PBE reference) $\Gamma$-X and X-X band gap errors are both only 0.04 eV. Furthermore, since the long-range electron correlation is mostly captured by the CCGF solver within ibDET, the starting-point dependence is significantly reduced from 0.71 eV to 0.08 eV ($\Gamma$-X gap) when using PBE vs. HF reference, which is also smaller than in full cell \GW+DMFT with similar embedding size (0.48 eV, $N_\mathrm{emb}=170$). 

ibDET also predicts accurate photoemission spectrum on 2D BN (Fig.~\ref{fig:bn_si}b). Previous full cell HF+DMFT simulation with a BN unit cell as the impurity yields accurate band gaps, but the spectrum shape shows some discrepancies, especially in the valence part~\cite{zhuEfficientFormulationInitio2020}, an indication of broken translational symmetry. In contrast, the density of states (DOS) predicted by HF+ibDET (200 orbitals in each embedding space) is in good agreement with full EOM-CCSD, suggesting the treatment of non-local electron correlation is substantially improved and the translational symmetry is preserved. The valence spectrum is near perfect, although the band gap is overestimated due to the large error in HF. \GW+ibDET (HF reference) further improves over HF+ibDET and achieves quantitative agreement with EOM-CCSD over a wide frequency range. 

\begin{figure}[hbt!]
\includegraphics[width=0.42\textwidth]{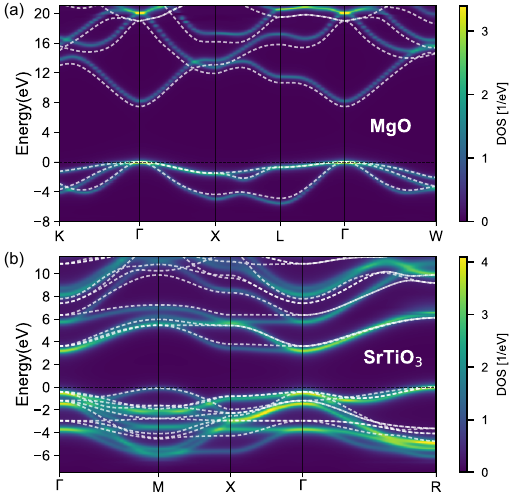}
\caption{Band structure of MgO and SrTiO\textsubscript{3} computed by \GW+ibDET (heat map) and $G_0W_0$@PBE (white dashes).}
\label{fig:mgosto}
\end{figure}

We then apply \GW+ibDET to study two metal oxides (MgO and SrTiO\textsubscript{3}) with large \textbf{k}-point sampling ($6 \times 6 \times 6$) impossible for standard EOM-CCSD implementation. MgO has an experimental band gap of 7.98$\sim$8.19 eV~\cite{voPerformancePeriodicEOMCCSD2024,renAllelectronPeriodicImplementation2021}, 
but \gw@PBE underestimates the band gap (7.43 eV\cite{zhuAllElectronGaussianBasedG0W02021}) and quasiparticle self-consistent \GW~largely overestimates (9.33 eV\cite{leiGaussianbasedQuasiparticleSelfconsistent2022}).
We performed \GW+ibDET (PBE reference) calculation using all-electron cc-pVTZ basis set, with 230 orbitals in each embedding space (2\% of $N_\mathrm{tot}$). As presented in Fig.~\ref{fig:mgosto}a and Table S6, \GW+ibDET greatly improves over \gw@PBE and predicts the band gap to be 8.22 eV, which is also consistent with recent EOM-CCSD benchmark (8.34 eV\cite{voPerformancePeriodicEOMCCSD2024}).

For the moderately correlated insulator SrTiO\textsubscript{3}, the experimental indirect band gap is 3.25 eV~\cite{vanbenthemBulkElectronicStructure2001}. Although SrTiO\textsubscript{3} has no open-shell $3d$ electrons, its lowest conduction bands are dominated by localized Ti-$3d$ orbitals, causing severe underestimation of the band gap by PBE (1.82 eV). \gw@PBE overestimates the band gap (3.62 eV), while various self-consistent \GW~schemes yield even larger overestimation errors~\cite{kangInfluenceWavefunctionUpdates2015,bhandariAllelectronQuasiparticleSelfconsistent2018}. We conducted \GW+ibDET calculations using all-electron def2-TZVP/def2-SVP basis sets~\cite{weigendBalancedBasisSets2005} for Ti and O, and GTH-DZVP-MOLOPT-SR/GTH-HF-rev basis/pseudopotential for Sr. As seen in Fig.~\ref{fig:mgosto}b and Table S7, with around 210 orbitals in each embedding space (1\% of $N_\mathrm{tot}$), \GW+ibDET (PBE reference) predicts the R-$\Gamma$ and $\Gamma$-$\Gamma$ band gaps to be 3.24 eV and 3.74 eV, in excellent agreement with experimental values. 
Comparing the \gw@PBE and \GW+ibDET band structures, we find that \GW+ibDET predicts broader valence band spectrum due to the shift of O-dominant peaks by 1$\sim$2 eV.

\onecolumngrid
\begin{figure*}[hbt!]
\includegraphics[width=\textwidth]{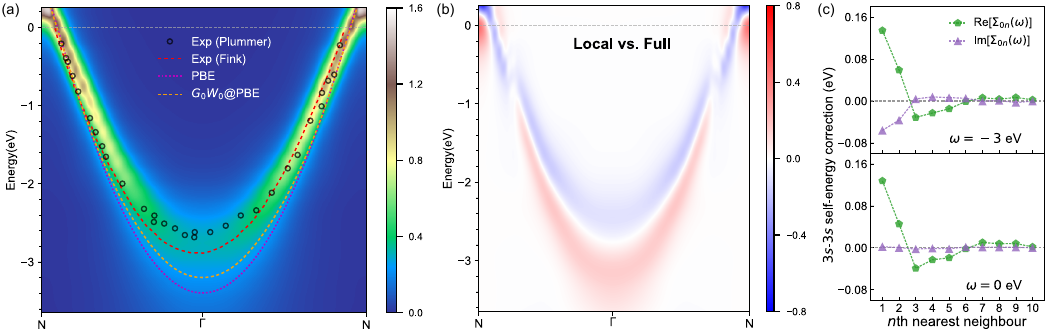}
\caption{
\GW+ibDET results for metallic sodium.
(a) Band structure computed by \GW+ibDET (heat map), compared against PBE, \gw@PBE, and ARPES experiments by Plummer~\cite{lyoQuasiparticleBandStructure1988} and Fink~\cite{potorochinLifetimeQuasiparticlesNearly2022}.
(b) 
DOS(``Local'') $-$ DOS(``Full''), computed by applying self-energy correction ($\Sigma^\mathrm{CC} - \Sigma^\GW$) to the full system (``Full'') or only to the diagonal block within each Na atom (``Local'').
(c) $3s$-$3s$ non-local self-energy correction ($\Sigma^\mathrm{CC} - \Sigma^\GW$) between Na atoms as the Na-Na distance increases.}
\label{fig:na}
\end{figure*}
\twocolumngrid

Finally, we demonstrate the applicability of ibDET to metallic systems, using sodium (Na) as an example. Although Na is usually considered as near-free-electron weakly correlated, DFT with LDA or GGA functionals severely overestimates the occupied bandwidth of Na (e.g., 3.41 eV from PBE) compared to those measured by angle-resolved photoemission spectroscopy (ARPES) experiments (2.65$\sim$2.78 eV)~\cite{lyoQuasiparticleBandStructure1988,potorochinLifetimeQuasiparticlesNearly2022}, which leads to long-standing debate over the nature of electron correlation in Na~\cite{mcclainSpectralFunctionsUniform2016,nilssonMultitierSelfconsistentGW2017,mandalElectronicCorrelationNearly2022,zhouCumulantGreenFunction2018,cracoLDA2019,Lischner2014}. Adding non-local static exchange in hybrid functionals yields even worse results~\cite{mandalElectronicCorrelationNearly2022}. 
The \GW~approximation is also insufficient, as \gw~only slightly improves over LDA and GGA (e.g., \gw@PBE value is 3.20 eV). Single-site DFT+DMFT and self-consistent \GW+EDMFT have been applied to this problem, where the impurity is a single Na-$3s$ orbital. LDA+eDMFT predicted a bandwidth of 2.84 eV~\cite{mandalElectronicCorrelationNearly2022}, which resulted in the conclusion that only local electron correlation within single Na atom needs to be captured beyond DFT. However, Ref.~\cite{nilssonMultitierSelfconsistentGW2017} showed contradictory result (3.2$\sim$3.3 eV, no improvement over \GW) from \GW+EDMFT, which suggested treating Na-Na non-local correlation beyond \GW~is important. Such discrepancy is likely due to the use of different effective interaction parameters within the downfolding scheme.

We thus apply ibDET, which is free of downfolding parameters and treats significantly larger embedding space (225 orbitals), to address this puzzle. The CCGF solver was previously shown to agree well with DMRG on the spectral function of a small uniform electron gas model at the relevant Wigner-Seitz radius $r_s=4$~\cite{mcclainSpectralFunctionsUniform2016}. Our \GW+ibDET (PBE reference) simulation employed GTH-cc-pVTZ basis set~\cite{yeCorrelationConsistentGaussianBasis2022} and GTH-HF-rev pseudopotential and $8 \times 8 \times 8$ \textbf{k}-mesh. 
In Fig.~\ref{fig:na}a, we find that \GW+ibDET achieves excellent agreement with the ARPES spectra~\cite{lyoQuasiparticleBandStructure1988,potorochinLifetimeQuasiparticlesNearly2022} and predicts an occupied bandwidth of 2.84 eV, 
significantly better than \gw@PBE and PBE.

Now that we have established the accuracy of \GW+ibDET, we further analyze the nature of electron correlation in metallic sodium. Specifically, we ask if the same good bandwidth prediction can be obtained with only local self-energy approximation, by limiting the $\Sigma^\mathrm{CC}-\Sigma^\GW$ self-energy correction to the diagonal block within each Na atom (this approximation is similar to single-site DFT+DMFT and \GW+EDMFT). In Fig.~\ref{fig:na}b, we find that, without non-local inter-atomic self-energy correction beyond \GW, the bandwidth predicted by ``local'' \GW+ibDET is 3.11 eV, only slightly improved over \gw@PBE (3.20 eV) and much worse than that predicted by full \GW+ibDET (2.84 eV). 
Furthermore, \GW+ibDET allows us to quantify the magnitude of real-space long-range electron correlation. In Fig.~\ref{fig:na}c, 
we find that the real part of inter-atomic $3s$-$3s$ self-energy correction ($\Sigma^\mathrm{CC}-\Sigma^\GW$) does not decay to zero until 6th nearest neighbour in distance, indicating the electron correlation is quite delocalized in metallic sodium. Thus, to quantitatively simulate spectral properties of sodium, we argue it is crucial to account for long-range electron correlation at a many-body level beyond DFT and \GW, as seen in \GW+ibDET.

\textit{Conclusion.} We have developed a new Green's function embedding formulation, interacting-bath dynamical embedding theory, for capturing local and non-local electron correlations on an equal footing in many-body simulation of solids. 
The main strength of this method is that it avoids uncontrolled errors associated with small impurity subspace and empirical truncations, while fully leveraging the power of advanced quantum chemistry solvers for treating long-range electron correlation effects. We have demonstrated that the \GW+ibDET approach achieves quantitative description of spectral properties across a wide range of materials and preserves the translational invariance. 
In particular, ibDET provides a capability to examine the effect of non-local (and even long-range) electron correlation in determining material-specific electronic properties. 
\GW+ibDET is thus a promising tool for tackling material problems in which non-local electron correlation plays a significant role. 

\begin{acknowledgments}
This work was supported by the National Science Foundation under Grant No.~CHE-2337991 and a start-up fund from Yale University. J.L. also acknowledges support from the Tony Massini Postdoctoral Fellowship in Data Science from Yale University.
\end{acknowledgments}

\bibliography{ref}

\end{document}